\documentclass[aps,prl,twocolumn,superscriptaddress,amsmath,amssymb]{revtex4-1}
\usepackage{graphicx}
\usepackage{float}
\usepackage{caption} 
\usepackage{bm} 
\usepackage{verbatim}  
\usepackage{mathrsfs}
\usepackage{xcolor}
\newcommand{\vect}[1]{\boldsymbol{\mathbf{#1}}}

\def\trD{\mathcal{D}}

\usepackage{dcolumn}  
\usepackage[caption=false]{subfig}
\usepackage{array}

\begin{document}
\title{Dry active turbulence in microtubule-motor mixtures}

\author{Ivan Maryshev}
\affiliation{Centre for Synthetic and Systems Biology, Institute of Cell Biology, School of Biological Sciences, University of Edinburgh,
Max Born Crescent, Edinburgh EH9 3BF, United Kingdom}

\author{Andrew B. Goryachev}
\affiliation{Centre for Synthetic and Systems Biology, Institute of Cell Biology, School of Biological Sciences, University of Edinburgh,
Max Born Crescent, Edinburgh EH9 3BF, United Kingdom}

\author{Davide Marenduzzo}
\affiliation{SUPA, School of Physics and Astronomy, The University of Edinburgh, James Clerk Maxwell Building, Peter Guthrie Tait Road, Edinburgh, EH9 3FD, United Kingdom}

\author{Alexander Morozov}
\affiliation{SUPA, School of Physics and Astronomy, The University of Edinburgh, James Clerk Maxwell Building, Peter Guthrie Tait Road, Edinburgh, EH9 3FD, United Kingdom}

\date{\today}
\begin{abstract}
We study the dynamics and phase behaviour of a dry suspension of microtubules and molecular motors. We obtain a set of continuum equations by rigorously coarse graining a microscopic model where motor-induced interactions lead to parallel or antiparallel ordering. Through numerical simulations, we show that this model generically creates either stable stripes, or a never-settling pattern where stripes periodically form, rotate and then split up. We derive a minimal model which displays the same instability as the full model, and clarifies the underlying physical mechanism. The necessary ingredients are an extensile flux arising from microtubule sliding and an interfacial torque favouring ordering along density gradients. We argue that our minimal model unifies various previous observations of chaotic behaviour in dry active matter into a general universality class.
\end{abstract}

\maketitle

Recent studies of active matter, comprising particles that convert internal energy to relative motion -- exerting force or torque dipoles on the surrounding medium as they do so -- reveal that these systems generally function far from equilibrium and possess no passive analogues~\cite{Marchetti2013}. Instead, their microscopic models can sometimes be grouped into ``universality'' classes, and identifying the corresponding equations is currently an area of active research~\cite{Marchetti2013,Toner1998,Toner2005,Stenhammar2013,Tiribocchi2015,Tjhung2018}. For systems with orientational order (i.e., active liquid crystals), two important classes of  models that emerged in the process are momentum-conserving (``wet'') incompressible systems~\cite{Simha2002,Marchetti2013} and non-momentum conserving (``dry'') compressible ones~\cite{Peshkov2012,Marchetti2013,Putzig2016,Srivastava2016}, with the vast majority of work dedicated to the former class. 

Here we study an example of dry active matter and consider the dynamics of pattern formation in mixtures of microtubules (MTs) and molecular motors (MMs)~\cite{Needleman2017, Marchetti2013}. These systems are relevant to both biological and synthetic instances of active matter. On the one hand, they incorporate the essential ingredients of the mitotic spindle~\cite{Mogilner2010,Burbank2007,Brugues2014}, on the other hand, they closely mirror the so-called ``hierarchical active matter'', which can be self-assembled in the lab from MTs and MMs, in the presence of polyethylene glycol~\cite{Sanchez2011,Sanchez2012,Guillamat2016}. 

Whilst the continuous description of the overdamped active biofilaments can be postulated on symmetry grounds~\cite{YounLee2001,Putzig2016}, it can also be derived by rigorously coarse-graining a specific underlying microscopic model~\cite{Kruse2000,Liverpool2003,Aranson2005, Ziebert2005,Aranson2006,Johann2016,Maryshev2018,Bertin2015}. 
This avenue is useful as it allows one to determine the effective parameters of the continuum theory in terms of the geometrical and physical quantities appearing in the microscopic model. Here, we follow this approach to describe a two-dimensional suspension of MTs interacting with kinesin-5-like MM~\cite{Kapitein2005}. 

Wet incompressible active gels are generically unstable to orientational fluctuations ultimately resulting in ``active turbulence''~\cite{Giomi2015,Wensink2012}. Here, we show that compressible dry MT-MM mixtures undergo seemingly similar chaotic dynamics, which we name \emph{dry active turbulence}; the underlying mechanism is, however, completely different. We derive a simple set of continuum equations that allows us to elucidate such a mechanism. We argue that dry active turbulence may constitute a general universality class shared between nematically ordered microtubules and flocking self-propelled particles.

We treat MTs as rigid rods of fixed length $l$ with distinct ends, denoted as ``$+$'' and ``$-$'', and consider ``$+$''-directed MMs, which are described by their distribution along individual MTs, as detailed below. Unlike in MT motility assays~\cite{Sumino2012} (where MTs are self-propelled), in MT-MM mixtures microtubular rods possess no constant velocity. Instead, filaments can only change position and orientation due to either thermal diffusion, or motor-mediated interactions.

\begin{figure}[H]
\centering
\includegraphics[width=0.9\linewidth]{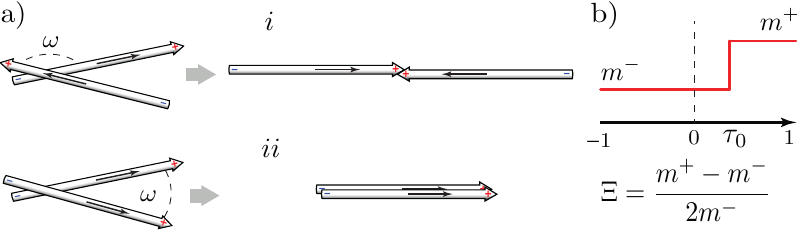}
\caption{(a) Collision rule including MT sliding 
(\emph{i}) and MT clustering (\emph{ii}), according to the incidence angle.  (b) Steady-state motor distribution considered in the anisotropic case (with inhomogeneous MT coverage by ``$+$''-directed motors).}
\label{fig1}
\end{figure}

The nature of motor-mediated interactions between filaments depends on the underlying microscopic detail. For instance, it was demonstrated that MMs able to associate with two filaments simultaneously can either cluster MTs, or actively separate them, depending on the initial configuration (Fig.~1). 
Importantly, our dynamics captures both possible outcomes -- consistent with the current view of most kinesin motors \cite{Kapitein2005,Fink2009}, and unlike previous work~\cite{Maryshev2018}, which solely focused on the case of polar clustering. Specifically, our interaction rule is the following. If the initial relative angle between rods exceeds some critical value (in our case $\pi/2$) then MTs first align in an anti-parallel way and then slide apart. 
Otherwise, MTs cluster to acquire the same position and orientation (Fig.~\ref{fig1}a). 

Within our model, MTs are covered by a steady-state static distribution of motors. Motor coverage may either be homogeneous or inhomogeneous~\cite{Leduc2012,Parmeggiani2003}, and is parametrized by two geometrical quantities: $\Xi=\left(m^+-m^-\right)/(2m^-)$, and $\tau_0$ (Fig.~\ref{fig1}b). Hereafter, we refer to $\Xi=0$ and $\Xi\ne 0$ as the isotropic and anisotropic cases, respectively.

In what follows, we work in two-dimensional Cartesian coordinates. Assuming that motor-induced rearrangements of MTs are fast with respect to diffusion, we treat them as instantaneous collisions. 
The probability distribution function, $P(\mathbf r,\phi)$, for a MT to be at a position $\mathbf r$ with an orientation $\mathbf n =(\cos\phi,\sin\phi)$, given by the angle $\phi$, 
obeys the following Boltzmann-like kinetic equation 
\begin{align}
&\qquad\partial_t P(\mathbf r,\phi)= D_{r}\partial_{\phi}^2P(\mathbf{r},\phi)+ \partial_i D_{ij}\partial_j P(\mathbf{r},\phi) \nonumber \\
&\qquad +\int d\vect{\xi}\Bigg[\int_{-\frac{\pi}{2}}^{\frac{\pi}{2}}d\omega W^+_1
P\left(\mathbf r_1,\phi_1\right)P\left(\mathbf r_2,\phi_2\right)\nonumber \\
&+\hspace{-1ex}\int\limits_{\pi/2}^{3\pi/2}\hspace{-1ex}
d\omega W^+_2
P\!\left(
	\mathbf r_1 \!+\! \frac{\eta l \mathbf n}{2},
    \phi_1      \!+\! \frac{\pi}{2}
\right)
P\!\left(
	\mathbf r_2 \!+\! \frac{\eta l\mathbf n}{2},
    \phi_2      \!+\! \frac{\pi}{2}
\right)\!
\Bigg]\nonumber \\
&\qquad
-\int d\vect{\xi}\int_{0}^{2\pi}d\omega W^-
P \left(
	\vect{r},\phi
\right) P \left(
	\vect{r}-\vect{\xi},\phi-\omega
\right).
\label{ME}
\end{align}
The first two terms on the r.h.s. of Eq.~\eqref{ME} represent contributions from diffusion, where $D_{r}$ is the rotational diffusion coefficient and $D_{ij}$ are components of the translational diffusion tensor~\cite{DoiEdwards,Maryshev2018}. The rest of the equation encodes our collision rules between MTs, and includes clustering and sliding, see Figure~\ref{fig1}a. The positions of the colliding MTs are given by $\mathbf r_{1,2}=\vect{r}\mp\frac{\vect{\xi}}{2}$, while their orientations are defined by the angles $\phi_{1,2}=\phi \mp \frac{\omega}{2}$; $\vect\xi$ and $\omega$ parametrise separations between MT centres and their orientations, respectively. The parameter $\eta$ determines the final relative displacement of MTs after sliding -- henceforth we consider $\eta=1$, corresponding to full separation. 
For needle-like MTs considered here, the collision rates $W^+_1$, $W^+_2$, $W^-$ only differ from zero 
when two MTs intersect in 2D; see~\cite{SI,Maryshev2018} for their explicit dependence on $\vect\xi$, 
$\omega$, $\Xi$, and $\tau_0$. 

\begin{figure*}[t]
\centering
\includegraphics[width=0.9\textwidth]{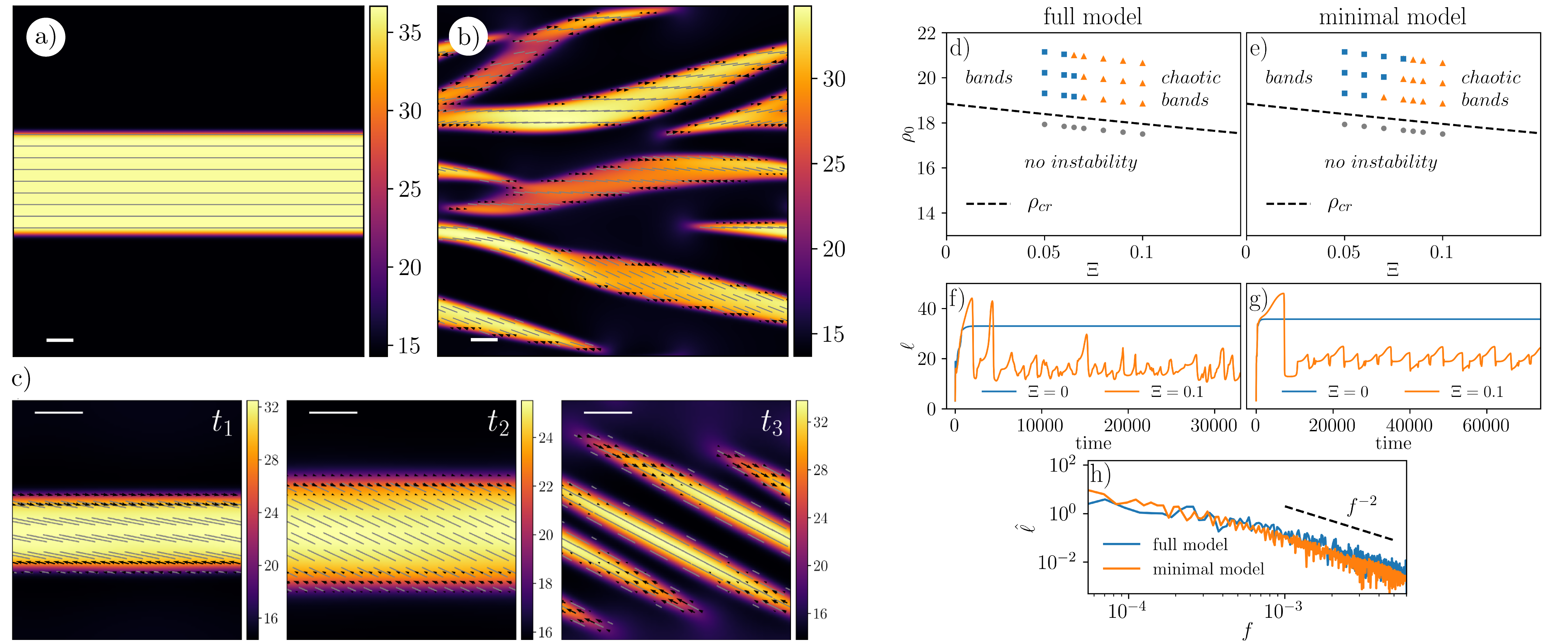}
\caption{ (a-c) Numerical simulations of the full model. (a) Formation of a stable stripe in the isotropic case ($\Xi=0,\rho_0=1.1\rho_{cr}$, system size $L=300$). (b) Chaotic dynamics for $\Xi\ne 0$ ($\Xi=0.1,\tau_0=0.5,\rho_0=1.1\rho_{cr}$, $L=300$). (c) Same as (b), but for $L=100$;  snapshots $t_1-t_2-t_3$ show the evolution of a nematic band. In (a-c) colormaps represent the MT density, black arrows denote the polar order field, and gray segments illustrate the largest eigenvector of the nematic alignment tensor $Q_{ij}$. Scale bar: $10$ $l$. (d,e) Phase diagram for the full (d) and minimal (e) model. Note that $\rho_N$ is much above the density range plotted~\cite{SI}. (f,g) Domain size $\ell$ versus time for the full (f) and minimal (g) model. (h) Fourier transform of $\ell$ versus frequency, for the full and minimal model.}
\label{fig2new}
\end{figure*}

We proceed by applying a rigorous coarse-graining procedure developed in~\cite{Maryshev2018} to Eq.~\eqref{ME} to derive a system of mean-field equations for the following fields: (i) the density of filaments $\rho$, (ii) their mean orientation $p_i$, and (iii) a tensorial field $Q_{ij}$ quantifying the nematic (apolar) ordering of MTs. 
These variables are defined as the first three moments of $P(\mathbf r,\phi)$:
\begin{gather}\label{moments}
\rho(\vect{r}) = \int_0^{2\pi}\!P(\vect r,\phi)d\phi, \qquad p_i(\vect{r})= \frac{1}{2\pi}\int_0^{2\pi}\! n_i\, P(\vect r,\phi)d\phi,\nonumber\\
Q_{ij}(\vect r) =\frac{1}{\pi}\int_{0}^{2\pi}\!\left(n_in_j-\frac{1}{2}\delta_{ij}\right) P(\vect r,\phi)d\phi,
\end{gather}
where $i,j=\{x,y\}$ denote the Cartesian components, and we introduced dimensionless units \cite{SI}.
The resulting equations contain a very large number of terms, as is often the case with kinetic theories, and their explicit form is given in~\cite{SI}. To study the dynamics predicted by this approach, we perform numerical simulations of the hydrodynamic equations and discuss representative results below (see Fig.~\ref{fig2new}.) Simulations are initialised from an isotropic uniform MT suspension with overall density $\rho_0$ and a small amount of noise. Without loss of generality, we set $\tau_0=1/2$, and vary $\Xi$ and $\rho_0$.

A linear stability analysis~\cite{SI} shows that the uniform isotropic state is linearly unstable towards the emergence of a globally-ordered nematic state, when $\rho_0>\rho_{cr}=6\pi/(1+\Xi(1-\tau_0))$. 
Additionally, for $\rho_{cr}<\rho_0<\rho_N$, this nematic state is itself unstable.  Simulations demonstrate that the latter instability leads to co-existence between high-density, nematically-ordered elongated domains and a low-density isotropic background (Fig.~2a and Suppl.~Movie 1). The outcome of this phase separation at late times depends on the value of the anisotropy parameter, $\Xi$. For small $\Xi$, domains coarsen to leave a single static band, whose size scales with the system size (Fig.~\ref{fig2new}a). Inside the band, MTs are ordered nematically, with residual polar order confined at the interface with the isotropic phase.
For large enough $\Xi$, we instead observe an ever-evolving pattern (Fig.\ref{fig2new}b and Suppl.~Movie 2), superficially reminiscent of ``active turbulence'' \cite{Giomi2010} in wet active gels. To characterise the properties of this spatiotemporal pattern, which we call \emph{dry active turbulence}, we plot the time evolution of the domain size, computed via the first moment of the structure factor~\cite{SI}, and its Fourier transform (Figs.~\ref{fig2new}f and h respectively). 
It is apparent that there is a selected lengthscale in the isotropic case, while the dynamics in the anisotropic case appear to be chaotic (as the Fourier transform in Fig.~\ref{fig2new}h contains all frequencies). Our findings are summarised in the phase diagram in Figure~\ref{fig2new}d. 

The kinetic pathway associated with dry active turbulence becomes apparent in simulations with smaller domains (Fig.~\ref{fig2new}c, Suppl.~Movie 3). These shows that the self-assembled nematically ordered MT bands undergo a cyclic process where they stretch perpendicular to their long direction, rotate, stretch and split, to reform later on. This process is quasi-periodic in smaller system, but appears to be chaotic in larger ones. 

To identify the fundamental mechanism leading to pattern formation in our system, we now search for a minimal model. We define the latter as a set of simple equations, which simultaneously satisfies two conditions. First, it needs to have qualitatively similar dynamics as the full model (Figs.~\ref{fig2new}a and b): it should retain both a transition between a uniform and a phase separated nematic state, as well as a regime with chaotic dynamics; in small domains, it should exhibit features similar to  Figure~\ref{fig2new}c. Second, we require that the location of the phase boundaries in the minimal and full models (Figs.\ref{fig2new}d and e), is quantitatively similar. As a first step, we exploit the observation that polar order plays a minor role (Fig.~\ref{fig2new}c), and adiabatically eliminate $p_i$ in favour of $\partial_i \rho$ and $\partial_jQ_{ij}$, keeping only the lowest order terms in spatial gradients (as in a hydrodynamic expansion~\cite{Wolf2004}). Then, we systematically switch off each term individually in the resulting equations, and compute the phase diagram; the term is only reinstated if its exclusion leads to a substantial change in the phase boundary location. 

This procedure yields the following dynamical equations for $Q_{ij}$ and $\rho$,
\begin{align}
\partial_{t}\rho = 
	& \nabla^2\left[\frac{1}{32}\rho+\mu\rho^2\right]
 	+ \partial_i\partial_j\left[\frac{\pi}{48} + \chi\rho\right] Q_{ij}\nonumber\\
    &-\lambda\nabla^2\left(Q_{kl}Q_{kl}\right), \label{MinModel1} \\ 
\partial_{t} Q_{ij} 
	=&  \left[4\left(\frac{\rho}{\rho_{cr}}-1\right)
    -\alpha Q_{kl}Q_{kl} +\kappa\nabla^2\right]Q_{ij}\nonumber\\
    & +\zeta\trD_{ij}\rho
 	  -\beta_1\trD_{ij}\left(Q_{kl}Q_{kl}\right)
 	  -\beta_2 Q_{kl}\trD_{ij}Q_{kl}              \label{MinModel2},
\end{align}
where we have introduced the operator $\trD_{ij}=\partial_i\partial_j-(1/2)\delta_{ij}\partial_k\partial_k$. 
The phase diagram corresponding to the minimal model is given in  Figure~\ref{fig2new}e.
All eight parameters in Eqs.~\eqref{MinModel1} and \eqref{MinModel2} -- $\mu$, $\chi$, $\lambda$, $\alpha$, $\kappa$, $\zeta$, $\nu$, $\beta$ -- are essential to get quantitative agreement with the full model; their expressions in terms of the microscopic quantities $\rho_0$, $\Xi$, $\tau_0$ and $\eta$ are given in~\cite{SI}. Within this set, $\zeta$ is the only parameter that can change sign -- the others are always positive. 

We now discuss the physical meaning of each term in Eqs.~\eqref{MinModel1} and \eqref{MinModel2}. First, $\mu$ and $\lambda$ determine the non-equilibrium chemical potential of our mixture: their main role is to set the values of the densities in the isotropic and nematic phases. Next, $\alpha$ is a non-equilibrium Landau coefficient setting the magnitude of order in the bulk (together with the term $4\left(\rho/\rho_{cr}-1\right)Q_{ij}$), while $\kappa$ is the nematic elastic constant. 
Similar terms are also present in a purely passive \emph{Model C}~\cite{HohenbergHalperin} describing, for instance, phase separation in passive liquid crystals. The key {\it qualitative} ingredients that produce chaotic behaviour in our model are the ``active'' terms proportional to $\chi$, $\zeta$, $\beta_1$ and $\beta_2$. Among them, $\chi$ is an ``extensile flux'', whose role is similar to that of an extensile stress in active gels~\cite{Simha2002,Marchetti2013}. This term enhances diffusion along the direction of the local nematic order (i.e., the eigenvector of  $Q_{ij}$ corresponding to its positive eigenvalue), and decreases it along the perpendicular direction. Second, $\zeta$ creates an effective torque at the interface, as the associated term depends on density gradients, which are largest at the interface. When $\zeta$ is positive (negative), it tends to orient MTs parallel (perpendicular) to an isotropic-nematic interface. Finally, $\beta_1$ and $\beta_2$ create modulation of the nematic ordering (i.e., the positive eigenvalue of $Q_{ij}$). 
These terms promote activity-induced disorder, and act similarly to a negative elastic constant in conventional liquid crystals. Additionally, they contribute to the interfacial torque at the boundary of a nematic band, where $Q_{kl}Q_{kl}$ drops sharply to zero, following the density field.

The minimal model is now simple enough for us to dissect the mechanisms underlying pattern formation. The kinetic pathway leading to non-equilibrium phase separation proceeds as follows. Starting from a uniform disordered solution with $\rho>\rho_{cr}$, MTs rapidly acquire orientational order, through the Landau coupling in Eq.\eqref{MinModel2}. At this point, the extensile active flux, arising from MT sliding, enhances diffusion along the nematic direction, and hinders it perpendicularly. When this effect is strong enough, the perpendicular diffusion becomes effectively negative, causing MT bundling and the formation of one or more nematically ordered high-density bands (see Fig.~\ref{fig2new} and  Suppl. Movies~1, 4). Notably, although the phase separation is driven by a non-equilibrium phenomenon (MM activity), the kinetic growth laws resemble canonical~\emph{Model C} phase separation in passive mixtures of liquid crystalline and isotropic fluids~\cite{HohenbergHalperin,Mata2014,SI}. 

Second, when 
$\Xi$ is sufficiently large, the $\beta_{1,2}$ terms dominate over both the restoring elastic constant $\kappa$ and the $\zeta$ term: the associated torque rotates the MTs at the nematic-isotropic interface, so that they tend to orient perpendicular to the band border. This interfacial alignment conflicts with the direction of the nematic order within the bulk of the band; it couples to the extensile flux to yield locally synchronous rotation (and stretching) of nematic bands as observed in our simulations. This cycle repeats, creating a never-settling pattern, as seen in our simulations in the dry active turbulent regime (Figs.\ref{fig2new} and 4a, and Suppl. Movies 3 and 5). As the sense of the emerging band rotation (clockwise or anticlockwise) is selected by spontaneous symmetry breaking, it may be different in different regions of our simulation domain, yielding a chaotic pattern (Fig.~4b, Suppl. Movies 2 and 6). 
Measuring the time evolution of the domain size in this regime yields statistically the same results as for the full model (Figs.~\ref{fig2new}g and h).

There is also a second mechanism that can destabilise a homogeneous nematic state, again dependent on $\beta_{1,2}$. A linear stability analysis starting from the uniform nematic phase~\cite{SI} shows that when these terms are large enough, they trigger the development of a modulation in $Q_{ij}$ -- in the direction parallel to that of the nematic order,  for $\beta_{1,2}>0$.
This instability is independent of density fluctuations and ultimately fragments the nematic phase into infinitely small microdomains.
This pathway to chaos is related to that identified in~\cite{Putzig2016,Srivastava2016} for dry active matter with near-uniform density. However, in our model this instability is only relevant for $\rho_0\gg \rho_{cr}$, and for lower $\rho_0$ is superseded by the turbulent phase separation dynamics discussed above.

While our minimal model is the result of a systematic coarse-graining, we can view Eqs.~\eqref{MinModel1} and \eqref{MinModel2}, more generally, as a phenomenological model that contains the lowest terms of the correct tensorial nature~\cite{BerisEdwards}. Upon coarse-graining, a microscopic model within the same universality class as the one studied here would, therefore, provide the expressions for the parameters in Eqs.~\eqref{MinModel1} and \eqref{MinModel2}, but would not generate extra terms. 
Indeed, setting $\beta_{1,2}=\lambda=\mu=0$ 
shows that our equations, in this limit, reduce to the minimal model for flocking of self-propelled particles with nematic order~\cite{Peshkov2012,Ngo2014,Shi2014,Grossmann2016}. We, therefore, propose Eqs.~\eqref{MinModel1} and \eqref{MinModel2} as a unifying model for dry active systems with nematic order. Recently, similar arguments were used to propose active versions of Models B and H~\cite{Stenhammar2013,Tiribocchi2015} in Hohenberg-Halperin classification~\cite{HohenbergHalperin}. We follow this analogy and refer to Eqs.~\eqref{MinModel1} and \eqref{MinModel2} as \emph{active Model C}. This model is in a different universality class with respect to active gel theory~\cite{Marchetti2013}, which exhibits instabilities in an active nematic fluid with constant density, whereas in our case patterns are always associated with a non-equilibrium phase separation. We want to stress that while previous work reported types of chaotic behaviour similar to the limiting cases of our model, either based on hydrodynamic~\cite{Peshkov2012,Ngo2014} or kinetic theories~\cite{Shi2014}, active model C unifies all this into a general universality class.

Analysis of active Model C with phenomenological coefficients re-enforces our physical interpretation of the instability modes. First, nematic-isotropic phase separation also occurs with $\zeta=\beta_{1,2}=0$, confirming that this phenomenon relies solely on a non-zero extensile flux, $\chi\ne 0$ (Fig.~3c, Suppl. Movie 7). Second, setting $\chi=0$ whilst retaining $\beta_{1,2}$ and $\zeta$ only leads to a uniform nematic phase, confirming that $\chi$ is necessary for any patterning (Suppl. Movie~8). Third, switching off only $\zeta$ leads to chaotic dynamics for a much wider parameter range, including $\Xi=0$ (Fig.~3d, Suppl. Movie~9), as now $\beta_{1,2}$ only need to compete with the elastic constant $\kappa$. Fourth, switching off only $\beta_{1,2}$ whilst retaining $\zeta>0$ does not lead to chaotic dynamics as in Fig.~2b and 3a,b, as there is no competition between the orientational order in the bulk and at the interface (see~\cite{SI}) . This case however,  yields another interesting instability associated with interfacial undulations and an elastic bend deformation in the nematic order (Fig.~3e, Suppl.~Movie 10). The ensuing patterns may also be chaotic for sufficiently large $\zeta$ (Suppl.~Movies 11, 12), and are similar to the structures seen experimentally in microtubule-kinesin mixtures~\cite{Sanchez2012}. 

For various values of its parameters, active Model C serves as a catalogue of patterns in dry active systems. As mentioned above, a sub-set of terms in Eqs.\eqref{MinModel1} and \eqref{MinModel2} was obtained in models of flocking of self-propelled particles with nematic order~\cite{Peshkov2012,Ngo2014,Shi2014,Grossmann2016}. Within those models, rigorous coarse-graining shows that $\chi$ and $\zeta$ are both positive, and, accordingly, the generic outcome found by numerical simulations~\cite{Peshkov2012,Ngo2014,Shi2014,Grossmann2016} is non-equilibrium phase separation and chaos through band undulations (as in Fig.~3e). Based on our phenomenological model and interpretation, we also expect dry active turbulence with contractile active flux, $\chi <0$, and interfacial torques favouring parallel alignment at the interface, as would occur when $\zeta$, or $\beta_{1,2}$ are positive. This scenario may be relevant for pattern formation in MT-MM mixtures where the underlying microscopic collision rules differ from those in Figure~1. 
Further work is required to identify the criteria for a microscopic model to belong to the universality class of our active Model C.

Discussions with Hugues Chat\'{e} are kindly acknowledged.
AG acknowledges funding from the Biotechnology and Biological Sciences Research Council of UK (BB/P01190X, BB/P006507). DM acknowledges support from ERC CoG 648050 (THREEDCELLPHYSICS). 

\begin{figure}
\includegraphics[width=1.\linewidth]{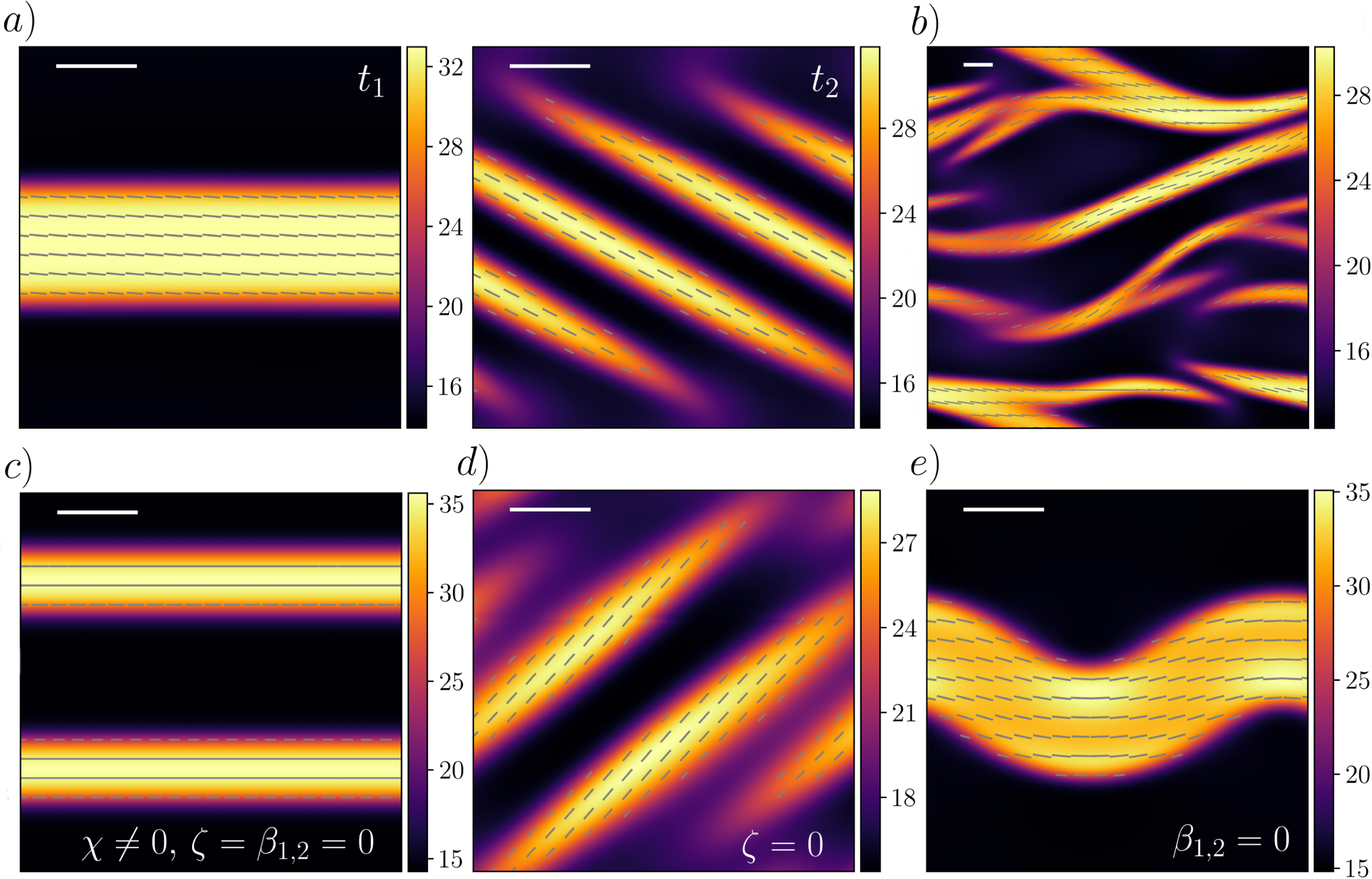}
\caption{Pattern formation within the minimal model. (a) Chaotic dynamics similar to Figure~2b ($\Xi=0.1,\tau_0=0.5$). (b) Chaotic dynamics for larger system size and anisotropy ($\Xi=0.3,\tau_0=0.5$). (c) Non-equilibrium phase separation with $\beta_{1,2}=\zeta=0$ ($\Xi=0$). (d) Chaotic dynamics with $\zeta=0$ ($\Xi=0$). (e) Interfacial undulation and chaos with $\beta_{1,2}=0$ ($\Xi=0$). For all plots $\rho_0=1.1\rho_{cr}$, scale bar: $10$ $l$.
}
\label{fig4}
\end{figure}

\enlargethispage{1\baselineskip}
\vspace{-2ex}

\clearpage

\renewcommand{\theequation}{S\arabic{equation}}
\renewcommand{\thefigure}{S\arabic{figure}}

\widetext
\begin{center}
\textbf{\large Dry active turbulence in microtubule-motor mixtures \\[2mm] Supplemental Material}
\end{center}

\begin{center}
Ivan Maryshev and Andrew B. Goryachev\\
\textit{\small Centre for Synthetic and Systems Biology, Institute of Cell Biology,
\\ School of Biological Sciences, University of Edinburgh,
\\ Max Born Crescent, Edinburgh EH9 3BF, United Kingdom}
\end{center}

\begin{center}
Davide Marenduzzo and Alexander Morozov\\
\textit{\small SUPA, School of Physics and Astronomy, The University of Edinburgh, 
\\ James Clerk Maxwell Building, Peter Guthrie Tait Road, Edinburgh, EH9 3FD, United Kingdom}
\end{center}

\section{Interaction rates}
\setcounter{equation}{0}
\setcounter{figure}{0}
\setcounter{table}{0}
\setcounter{page}{1}

The interaction functions $W$ defined in the main text determine the rates at which two microtubules (MTs) at $(\vect r_1,\phi_1)$ and $(\vect r_2,\phi_2)$ are displaced and reoriented by molecular motors (MMs). In our approach these rates have the following general form:
\begin{align}
&W\left(\vect{r}_1,\phi_1;\vect{r}_2,\phi_2\right) = G\,
\underbrace
{
\Theta\!\big(1 -\left|\tau_1\right|\big)
\Theta\!\big(1 -\left|\tau_2\right|\big)
}_{\text{probability of intersection}}
\underbrace
{
\bigl\{ 
1+\Xi \bigl[\Theta( \tau_1 - \tau_0)+\Theta(\tau_2 - \tau_0)\bigr]
\bigr\}
}_{\text{dependence on the local MM density}}
.\label{SKernel}
\end{align} 

The constant $G$ is proportional to the motor properties (e.g., their processivity and the overall density); it varies with the motor type and will be removed from the model by a rescaling, see below. The product of the Heaviside functions gives the geometric probability of two MT intersecting in 2D: since we assume MMs to be rods of negligible thickness, $W$ should be non-zero only when MTs intersect in their original configuration (in other words we do not consider long-range interactions). $\tau_{1,2}$ are the positions of the intersection point along the two MTs. We parametrise these position such that $\tau=0$ at the MT centre and $\tau=\pm 1$ corresponds to the ``+''/``-''-ends, respectively. The expression within the curly brackets depends on the local density of MMs at the intersection point and introduces anisotropy in the interaction, which arises only when $\Xi\neq0$. $\tau_{0}$ is the position of the interface between low and high MM density on a particular MT.  
The derivation of Eq.\eqref{SKernel} is provided in \cite{Maryshev2018}.

Eq.\eqref{SKernel} can be written in terms of $\xi$, $\psi$, $\phi$, and $\omega$ as
\begin{align}
W\left(\vect{r}_1,\phi_1;\vect{r}_2,\phi_2\right)
 =& G\,
\Theta\!\left(|\sin \omega| -\frac{2\xi}{l}\left|\phi_1-\psi\right|\right)
\Theta\!\left(|\sin \omega| -\frac{2\xi}{l}\left|\phi_2-\psi\right|\right)
\nonumber\\&
\times
  \left\{
  1+
  \Xi\left[
  	\Theta\left( \frac{2\xi}{l}\frac{\sin\left(\phi_1-\psi\right)}{\sin \omega}-\tau_0\right)
  	+
  	\Theta\left( \frac{2\xi}{l}\frac{\sin\left(\phi_2-\psi\right)}{\sin \omega}-\tau_0\right)
  	\right]
\right\},
\end{align}
where $\vect{\xi}=\xi\left(\cos\psi,\sin\psi\right)=\vect r _2 -\vect r_1$ is the separation vector between MT centres, and $\omega=\phi_2-\phi_1$ is the angle between their orientations; $l$ is the MT length.

The interaction rates used in the main text read:
\begin{align}
W^+_1&\equiv W\left(\mathbf r-\frac{\vect\xi}{2},\phi-\frac{\omega}{2};\mathbf r+\frac{\vect\xi}{2},\phi+\frac{\omega}{2}\right),\nonumber\\
W^+_2&\equiv W\left(\mathbf r-\frac{\vect\xi}{2}+\frac{\eta l \mathbf n}{2},\phi-\frac{\omega}{2}+\frac{\pi}{2};\mathbf r+\frac{\vect\xi}{2}+\frac{\eta l \mathbf n}{2},\phi+\frac{\omega}{2}+\frac{\pi}{2} \right),\nonumber\\
W^-&\equiv W\left(\mathbf r,\phi;\mathbf r -\vect\xi,\phi-\omega\right).
\end{align}

\section{Full Model}
Using the techniques from Ref.~[1], we coarse-grain our microscopic model to arrive at the following equations for the evolution of density ($\rho$), polar order ($p_i$), and nematic alignment tensor ($Q_{ij}$) (this set of equation is referred to as the "full model" in the main text):
\begin{align}
\partial_{t}\rho  = &\frac{1}{32}\nabla^2\rho
+\frac{\pi}{48}\partial_i\partial_j Q_{ij}
	+\left(1+a_3\right)\frac{\pi}{4}
	\left[
		-\frac{1}{12\pi^2}\nabla^2\rho^2+\frac{1}{9}\nabla^2Q_{ij}Q_{ij}-\frac{1}{9\pi}\partial_i\partial_j(\rho Q_{ij})
	\right]\nonumber\\&
	+\left(1+a_1\right)\frac{\eta^2\pi}{4}
	\left[\frac{1}{4\pi^2}\nabla^2\rho^2-\frac{1}{3}\nabla^2Q_{ij}Q_{ij}+\frac{1}{2\pi}\partial_i\partial_j(\rho Q_{ij})-2\partial_i\partial_j\left( p_i p_j\right)
	\right]-\frac{91}{69120\pi}\rho_0\nabla^4\rho, 
\label{FullRho}	
\\
\partial_{t}  p_i  =&  - p_i+\frac{5}{192}\nabla^2 p_i
+\frac{1}{96}\partial_i(\partial_k p_k)
	+\left(1+a_1\right)\left[
		- \frac{3}{11\pi}\rho p_i+\frac{29}{19} Q_{ij} p_j 
		- \frac{7}{18} A_3Q_{kl}Q_{kl} p_i 
	\right]
	\nonumber\\&
	+a_2
	\frac{1}{8}
	\Bigg[ 
		-\frac{1}{4\pi^2}\partial_i\rho^2
		+\left( 3 p_i(\partial_k p_k)
			+ ( p_k \partial_k)p_i
			- \frac{1}{3} \partial_i(p_kp_k)
		\right)
		\nonumber\\&\qquad\quad
		-
		\left(
		\frac{1}{12\pi}\partial_j(\rho Q_{ij})+\frac{1}{2\pi}Q_{ij} \partial_j \rho
		\right)
		+
		\left(
		\frac{113}{180}\partial_i(Q_{kl}Q_{kl})-\frac{53}{45}Q_{ij}\partial_k Q_{jk}
		\right)
	\Bigg]
	\nonumber\\&
	+\left(
		1+a_3
	\right)\frac{1}{720}
	\Bigg[
		-\frac{31}{ \pi }p_i\nabla^2\rho
		-\frac{6}{ \pi }p_k\partial_k\partial_i\rho
		+\frac{19}{ \pi }\left(\rho\nabla^2 p_i
		-\frac{1}{2 }\nabla^2(\rho p_i)\right)
		-61 p_i\partial_k \partial_l Q_{k l}
		\nonumber\\&\qquad\qquad
		-\frac{7}{ \pi }\left(
		(\partial_i\rho)(\partial_k p_k)
		- \rho\partial_i(\partial_k p_k)
		+\partial_i(p_k\partial_k\rho)\right)
		+49(p_k\partial_k)(\partial_j Q_{ij})
		+31 p_k \partial_i\partial_l Q_{k l}
		+10Q_{k l}\partial_i\partial_k  p_l
		\nonumber\\&\qquad\qquad
		+28Q_{il}\partial_l (\partial_k p_k)
		-19Q_{ik }\nabla^2 p_k 
		-14\partial_k \partial_l (Q_{k l} p_i)
		+9\partial_i\partial_k (Q_{k l} p_l)
		-\frac{9}{2}\nabla^2(Q_{ij}p_j)
	\Bigg]
	\nonumber\\&
	+\left(1+a_1\right)
	\frac{\eta}{2}
	\Bigg[ 
		\frac{1}{4\pi^2}\nabla\rho^2
		+\frac{1}{2\pi}\partial_j(\rho Q_{ij})
		-\frac{1}{3}\partial_iQ_{kl}Q_{kl}
		-2\partial_j(p_ip_j)
	\Bigg]
	\nonumber\\&
	+
	a_2
		\frac{\eta}{44}
		\bigg[
		\frac{9}{8 \pi }\left(
		2(p_k\partial_k)\partial_i\rho
		-2\rho\partial_i(\partial_kp_k)
		-\rho\nabla^2 p_i
		+p_i\nabla^2\rho\right)
		+2\left(p_i\partial_k \partial_l Q_{k l}
		-Q_{k l}\partial_k \partial_l p_i\right)
		\nonumber\\&\qquad\qquad\quad
		+\frac{5}{2}((p_k\partial_k)(\partial_j Q_{ij})
		-Q_{il }\partial_l (\partial_k p_k))
		+\frac{3}{2}(Q_{k l}\partial_i\partial_k  p_l
		- p_k \partial_i\partial_l Q_{k l})
	\bigg],
	\label{FullP}
\\
\partial_{t}Q_{ij}  = & -4Q_{ij}+\frac{1}{32}\nabla^2 Q_{ij}
+\frac{1}{192\pi}\trD_{ij}\rho
	+\left(1+a_1\right)
    \left[\frac{2}{3\pi}\rho Q_{ij}
    -A_4\frac{6}{5}(Q_{kl}Q_{kl})Q_{ij}\right]
	\nonumber\\&
	+a_2	
	\Bigg[\!
		-\frac{1}{16\pi}
        \left(
			\partial_i(\rho p_j)
            +\partial_j(\rho p_i)
            -\delta_{ij}\partial_k\left(\rho p_k\right)	
		\right)
		-\frac{\rho}{24\pi}\left(
			\partial_i p_j+\partial_j p_i
            -\delta_{ij}(\partial_kp_k )	
		\right)
		+\frac{1}{4}Q_{ij }(\partial_k p_k)
        +\frac{5}{12}(p_k\partial_k)Q_{ij}
	\Bigg]
	\nonumber\\&
	+\left(
		1+a_3
	\right)\frac{1}{8}
	\Bigg[
			-\frac{1}{36\pi^2}
			\!\left(
			3\left[
				(\partial_i\rho)(\partial_j\rho)-\frac{\delta_{ij}}{2}(\partial_k\rho)^2
			\right]
			+\rho\trD_{ij}\rho
			\right)
			+\frac{5}{9\pi}\rho\nabla^2Q_{ij}
			-\frac{1}{3\pi}Q_{ij}\nabla^2\rho
			-\frac{1}{6\pi}\nabla^2(\rho Q_{ij})
			\nonumber\\&\qquad\qquad\qquad
			-\frac{11}{45}Q_{ij}\partial_k \partial_l Q_{k l}
			-\frac{1}{6}\partial_k \partial_l (Q_{ij} Q_{k l})
			+\frac{7}{15} Q_{k l} \partial_k \partial_l Q_{ij}
			+\frac{1}{12}\trD_{ij}(Q_{kl}Q_{kl})
			-\frac{1}{5} Q_{k l}\trD_{ij}Q_{k l}
	\Bigg]
	\nonumber\\&
	+\left(1+a_1\right)\frac{\eta^2}{8}
	\Bigg[
	\frac{1}{4\pi^2}\trD_{ij}\rho^2
	-\partial_i\partial_j(p_kp_k)
	-2\nabla^2(p_ip_j)
	-\frac{5}{6}\trD_{ij}(Q_{kl}Q_{kl})
+\partial_k \partial_l(Q_{k l} Q_{ij})
	+\frac{1}{2\pi}\nabla^2(\rho Q_{ij})
	\Bigg].
	\label{FullQ}
\end{align}
These equations were rendered dimensionless by scaling time, space and the Fourier harmonics of $P$ by $D_r^{-1}$, $L$ and $G L^2/D_r$, respectively; the indices refer to the two-dimensional Cartesian components and the Einstein summation convention is employed; $\nabla^2=\partial_k\partial_k$, and we introduced the operator 
$\trD_{ij}=\partial_i\partial_j-\frac{\delta_{ij}}{2}\partial_k\partial_k$. Note, that these equations can be written in a more compact form in terms of complex fields and the Wirtinger derivatives $\underline \nabla=\partial_x+i\partial_y$ and $\underline \nabla^*=\partial_x-i\partial_y$, where $^*$ denotes complex conjugation. However, we find the resulting equations more difficult to read and prefer to keep the original notation. 

Coefficients $A_3$ and $A_4$ are coming from the adiabatic elimination of higher Fourier modes of $P(\mathbf{r},\phi)$; for the details of the closure procedure see [1]. Their expressions are given by: 
\begin{align}
	A_4&=3\left( \frac{1}{1+a_1} +\frac{ \rho _0}{5 \pi }\right)^{-1}\!,\quad
    A_3=\left(\frac{15 }{8}\frac{1}{1+a_1}+\frac{19 \rho _0}{48 \pi }\right)^{-1}\!.
\end{align}
Finally, we have also introduced the following quantities that depend on $\Xi$ and $\tau_0$:
\begin{align} 
a_1=\Xi\left(1-\tau_0\right),\qquad
a_2=\Xi\left(1-\tau_0^2\right),\qquad
a_3=\Xi\left(1-\tau_0\left(1+\tau_0^2\right)/2\right).
\end{align}

\section{Minimal Model}
As discussed in the main text, as a first step in deriving a minimal model, we use our observation that the polar order plays only a minor role in the simulations of the full model. By keeping only the lowest order terms in spatial gradients in Eq.~\eqref{FullP}, we can adiabatically eliminate $p_i$ from the other equations by replacing it with
\begin{align}
p_i = \frac{\rho_0}{1+(1+a_1)\frac{3}{11\pi} \rho_0}\left[ \left(
	\frac{1+a_1}{4\pi^2}\eta-\frac{a_2}{16\pi^2}
	\right) \partial_i \rho 
	+ \left(
	\frac{1+a_1}{4\pi}\eta-\frac{a_2}{96\pi}
	\right) \partial_j Q_{ij} 
	\right].
\end{align}
This procedure resulted in two dynamical equations for $\rho$ and $Q_{ij}$. We then systematically switched off each term individually in these equations, and computed the resulting phase diagram in each case. The term was reinstated only if it significantly changed the position of the phase boundary, as compared with the phase diagram of the full model. By following this procedure, we obtained the following minimal model 
\begin{align}
\partial_{t}\rho = 
	& \nabla^2\left[\frac{1}{32}\rho+\mu\rho^2\right]
 	+ \partial_i\partial_j\left[\frac{\pi}{48} + \chi\rho\right] Q_{ij}-\lambda\nabla^2\left(Q_{kl}Q_{kl}\right), \label{SMMinModel1} \\ 
\partial_{t} Q_{ij} 
	=&  \left[4\left(\rho/\rho_{cr}-1\right)
    -\alpha Q_{kl}Q_{kl} +\kappa\nabla^2\right]Q_{ij}+\zeta\trD_{ij}\rho
 	  -\beta_1\trD_{ij}\left(Q_{kl}Q_{kl}\right)
 	  -\beta_2 Q_{kl}\trD_{ij}Q_{kl}              \label{SMMinModel2},
\end{align}
where the parameters are given by

\noindent
\begin{minipage}{\textwidth}
\setlength{\jot}{1ex}
\begin{align}
\mu=&\frac{1+a_1}{16\pi}\eta^2-\frac{1+a_3}{48\pi},\quad
\chi=\frac{1+a_1}{8}\eta^2-\frac{1+a_3}{36},\quad
\lambda=\pi\left(\frac{1+a_1}{12}\eta^2-\frac{1+a_3}{36}\right),
\quad
\alpha=\frac{2(1+a_1)}{ \frac{5}{1+a_1} +\frac{ \rho _0}{ \pi }},
\nonumber\\
\kappa=&
 \frac{1}{32}
 +\left(
 \frac{1+a_1}{16\pi}\eta^2
 +\frac{7(1+a_3)}{144\pi} 
 \right)
 \rho_0
-\frac{5a_2}{48\pi\left(1+\left(1+a_1\right)\frac{3}{11\pi}\rho_0\right)}
	\left(
	\frac{1+a_1}{4\pi}\eta-\frac{a_2}{96\pi}
	\right)\rho_0^2,
\nonumber\\
\zeta=&\frac{1}{192\pi}
 +\left(
 \frac{1+a_1}{16\pi^2}\eta
 -\frac{1+a_3}{288\pi^2} 
 \right)\rho_0
 -\frac{5a_2}{24\pi\left(1+\left(1+a_1\right)\frac{3}{11\pi}\rho_0\right)}
	\left(
	\frac{1+a_1}{4\pi^2}\eta-\frac{a_2}{16\pi^2}
	\right)\rho_0^2,
\nonumber\\
\beta_1=&
\frac{5\eta^2\left(1+a_1\right)}{48}-\frac{1+a_3}{96},\quad\quad\
\beta_2=
\frac{1+a_3}{40},
\quad\quad
\rho_{cr} = \frac{6\pi}{1+a_1}.
\end{align}

We note here that the minimal equations presented above cannot be derived from the full model using the amplitude-equation-like techniques where only terms up to a particular order in the distance to an instability threshold are preserved. Application of such techniques to our full model results in a model with a smaller number of terms than the minimal one presented above, and, as has already been noted, all the terms in our minimal model are required to reproduce the phase behaviour of the full system of equations.

        \centering
        \includegraphics[width=0.99\textwidth]{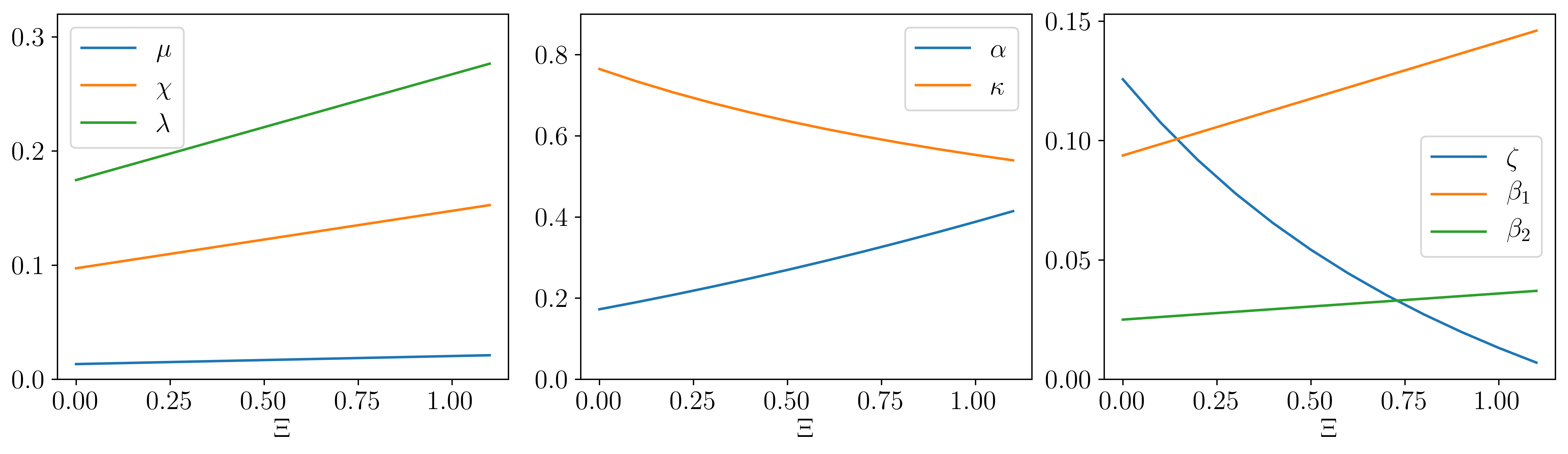}

        \vspace{-2ex}
        \captionof{figure}{Dependence of the minimal model parameters on $\Xi$ ($\rho_0=1.1\rho_{cr},\tau_0=0.5$).}
\end{minipage}

\section{Linear Stability Analysis}

Here we perform a linear stability analysis of the minimal model, Eqs.~\eqref{SMMinModel1} and \eqref{SMMinModel2}. Without loss of generality, we assume that the base state has a uniform density $\rho_0$ and a uniform nematic order of strength $Q_0$, oriented along the $x$-direction. We introduce infinitesimal perturbations to the $\rho$ and $Q_{ij}$ fields
\begin{align}
    \rho(\vect r,t)&=\rho_0+ \delta{\rho} e^{i \mathbf{k}\cdot \mathbf{r}}e^{\hat\sigma t},\nonumber\\
    Q_{xx}(\vect r,t)&=Q_0 + \delta{Q_{xx}} e^{i \mathbf{k}\cdot \mathbf{r}}e^{\hat\sigma t},\nonumber\\
    Q_{xy}(\vect r,t)&=\delta{Q_{xy}} e^{i \mathbf{k}\cdot \mathbf{r}}e^{\hat\sigma t},
\end{align}
where $k_x$ and $k_y$ set the lengthscale of the perturbation, and $\hat\sigma$ is a temporal eigenvalue. We substitute these expressions into Eqs.\eqref{SMMinModel1} and \eqref{SMMinModel2}, and linearise the resulting equations with respect to the perturbations to obtain
\begin{align}
\renewcommand*{\arraystretch}{1.5}
& \hat\sigma
\begin{pmatrix}
	\delta{\rho}\\\delta{Q_{xx}}\\\delta{Q_{xy}}
\end{pmatrix}
\!=\! \nonumber \\
& 
\begin{pmatrix}
-(1/32 +2\mu \rho _0) k^2-\chi Q_0 \bar{k}^2
&\,
-(\pi/48+\chi \rho _0 )\bar{k}^2 + 4  Q_0 \lambda k^2
&\,
-(\pi /24+2\chi \rho _0) k_x k_y
\\
4 Q_0/\rho_{cr} -\zeta \bar{k}^2/2
&\,
4\left(\rho_0/\rho_{cr}-1\right) - 6\alpha Q_0^2
+(2\beta_1+\beta_2)Q_0 \bar{k}^2
-\kappa k^2
&\,
0
\\
-\zeta k_x k_y
&\,
2 (2\beta_1+\beta_2)Q_0  k_x k_y
&\,
4\left(\rho_0/\rho_{cr}-1\right) - 2\alpha Q_0^2 - \kappa k^2
\end{pmatrix}\!\!\!
\begin{pmatrix}
	\delta{\rho}\\\delta{Q_{xx}}\\\delta{Q_{xy}}
\end{pmatrix}\!,
\label{LinStabQ}
\end{align}
where $k^2=k_x^2+k_y^2$, and $\bar{k}^2= k_x^2-k_y^2$. We proceed by studying the linear stability of various base states.

\subsection{Stability of the Homogeneous and Isotropic State}

Linear stability of the homogeneous and isotropic state is determined by the eigenvalue problem, Eq.\eqref{LinStabQ} with $Q_0=0$. Explicitly solving the eigenvalue problem, yields
\begin{align}
\hat\sigma = 4\left(\rho_0/\rho_{cr}-1\right) - \kappa k^2
\end{align}
for the most unstable eigenvalue. The instability sets in at $k=0$ and $\rho_0=\rho_{cr}$, corresponding to the transition to a globally-ordered nematic state (see Fig.~\ref{MinModPhaseD}\,a).

\subsection{Linear Stability of the Nematic State}

\begin{figure}[h!]%
    \centering
    \subfloat[Isotropic state]{{\includegraphics[width=0.45\textwidth]{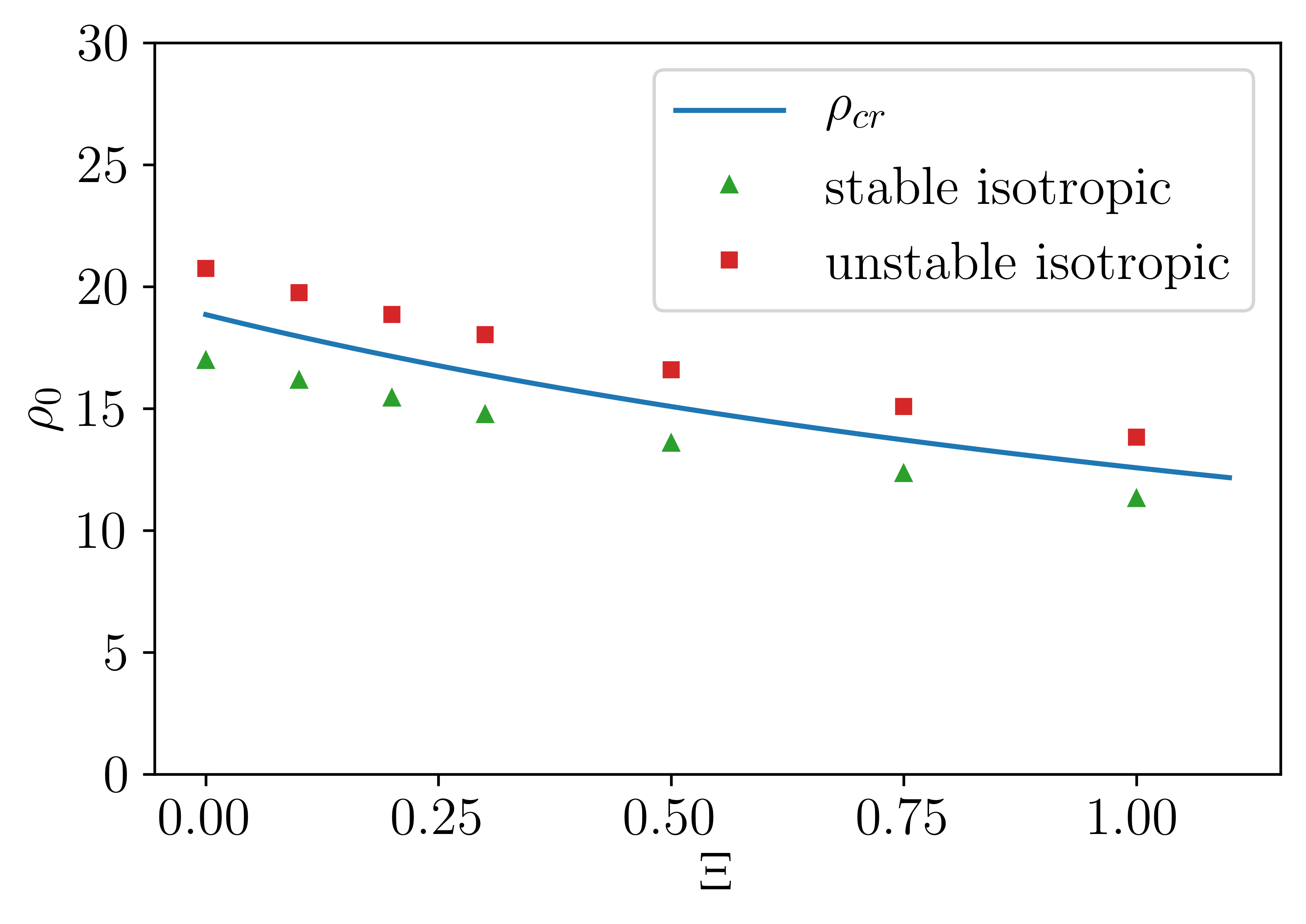}}}%
    \quad
    \subfloat[Nematic state]{{\includegraphics[width=0.46\textwidth]{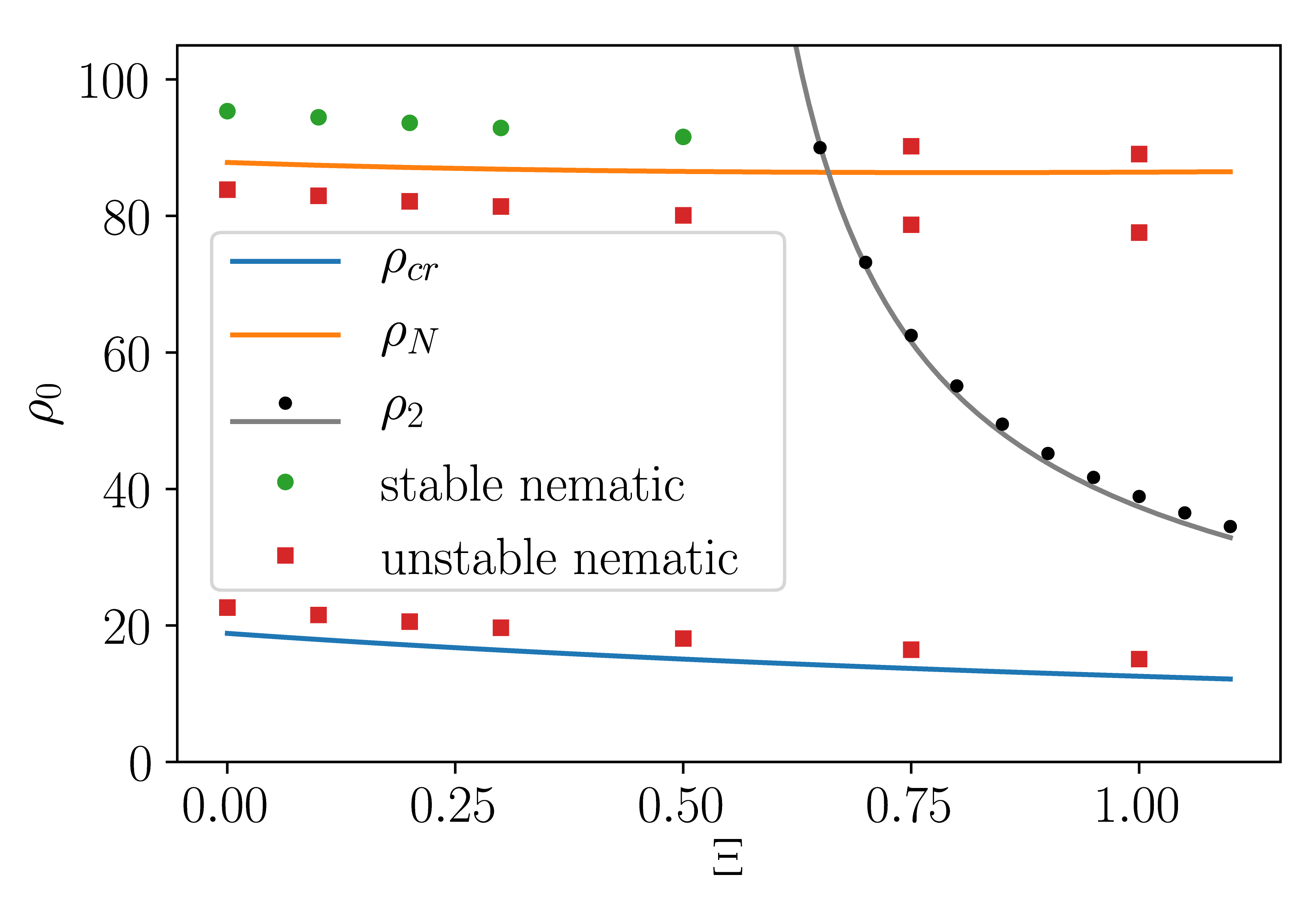} }}%
    \caption{Liner stability of the minimal model. (a) Stability of the homogeneous isotropic state, green triangles and red squares represent stable and unstable solutions obtained in the numerical simulations. (b) Stability of the nematic phase. The region between the blue and orange lines is unstable to phase separation; the region above the black dots denote the second instability described in the text (the gray line is the analytical approximation for this instability based on Eq.~\eqref{disprel2}). Green circles and red squares represent stable and unstable solutions obtained in the numerical simulations. For both cases $\eta=1$ and $\tau_0=0.5$.   
    }%
\label{MinModPhaseD}
\end{figure}

For $\rho_0>\rho_{cr}$, the homogeneous and isotropic state is unstable towards the formation of a global nematic phase with the amplitude $Q_0$, given by the spatially-independent terms in Eq.\eqref{SMMinModel2}
\begin{align}
Q_0=\sqrt{\frac{2}{\alpha}\left(\frac{\rho_0}{\rho_{cr}}-1\right)}.
\end{align}
Using this value in Eq.\eqref{LinStabQ} yields an eigenvalue problem that is too complicated to analyse analytically, and, instead, we study it numerically using Wolfram Mathematica. First, we observe that the globally oriented nematic state is always linearly unstable for $\rho_{cr}<\rho_0<\rho_N$ (i.e., the region between the blue and orange lines in Fig.~\ref{MinModPhaseD}\,b), where the upper phase boundary $\rho_N$ is determined numerically. 
The most unstable perturbations correspond to $k_x=0$, with the eigenvector in the form $\left(\delta\rho,\delta Q_{xx}, 0\right)$. This instability results in the modulation of the density and nematic order in the direction perpendicular to the nematic direction, and indicates the formation of the nematic bands, discussed in the main text. 

\subsection{Second Linear Instability of the Nematic State}

As discussed in the main text, for densities significantly larger that $\rho_{cr}$, there exists another linear instability of the global nematic state, which is different from the one discussed above. Numerical analysis shows that the corresponding eigenvector has a significant $\delta Q_{xx}$ component, and a very small density modulation $\delta \rho$. To get an insight into the nature of this instability, we set $\delta \rho$ to zero in Eq.\eqref{LinStabQ} to obtain a simple problem with the most unstable eigenvalue given by
\begin{align}
\hat\sigma = 4\left(\rho_0/\rho_{cr}-1\right) - 6\alpha Q_0^2
+(2\beta_1+\beta_2)Q_0 \bar{k}^2
-\kappa k^2.
\label{disprel2}
\end{align}
For all the values of parameters discussed in this work, the coefficient in front of $k_y^2$ is always negative, and we conclude that the most unstable eigenvalue corresponds to $k_y=0$. This eigenvalue becomes positive when $(2\beta_1+\beta_2)Q_0 > \kappa$. For the parameters used in our analysis, $\eta=1$ and $\tau_0=1/2$, this condition can be satisfied for $\Xi>0.49$, and the corresponding densities above which the instability arises are given in Fig.~\ref{MinModPhaseD}\,b as black circles (the analytical approximation for the instability boundary, Eq.~\eqref{disprel2}, is shown as a gray line). As Eq.~\eqref{disprel2} suggests, our minimal model does not predict a selected lengthscale for this instability due to the lack of higher-order spatial gradients in Eqs.~\eqref{SMMinModel1} and \eqref{SMMinModel2}, and, instead, the fastest growth is observed at the smallest scale available. This instability exists only for relatively large values of $\Xi$ and $\rho_0$ and is superseded by the main instability discussed above and in the main text.

\section{Coarsening}

As was mentioned in the main text, in the case where the interaction rates are isotropic ($\Xi=0$), nematic domains undergo a coarsening process and tend to form one band in steady state.

To characterise the way in which domains coarsen, we here quantify how the typical domain length scale $\ell$ grows with time. First, we compute the structure factor, $S= \langle\rho(t,\mathbf k)\rho(t, -\mathbf k)\rangle$, by averaging the output of the simulation at late times. Then, we define $\ell$ as 
\begin{align}
\ell(t)=2\pi \frac{\int S({\mathbf k},t) d{\mathbf k}}{\int k S({\mathbf k},t) d{\mathbf k}},
\end{align}
where $k=|{\mathbf k}|$.
Simulations to compute $\ell$ as a function of time $t$ are initialised with a system with uniform density and nematic order, with a small amount of noise.

\begin{figure}[h!]
    \centering
    \begin{minipage}{\textwidth}
        \centering
        \includegraphics[width=0.4\textwidth]{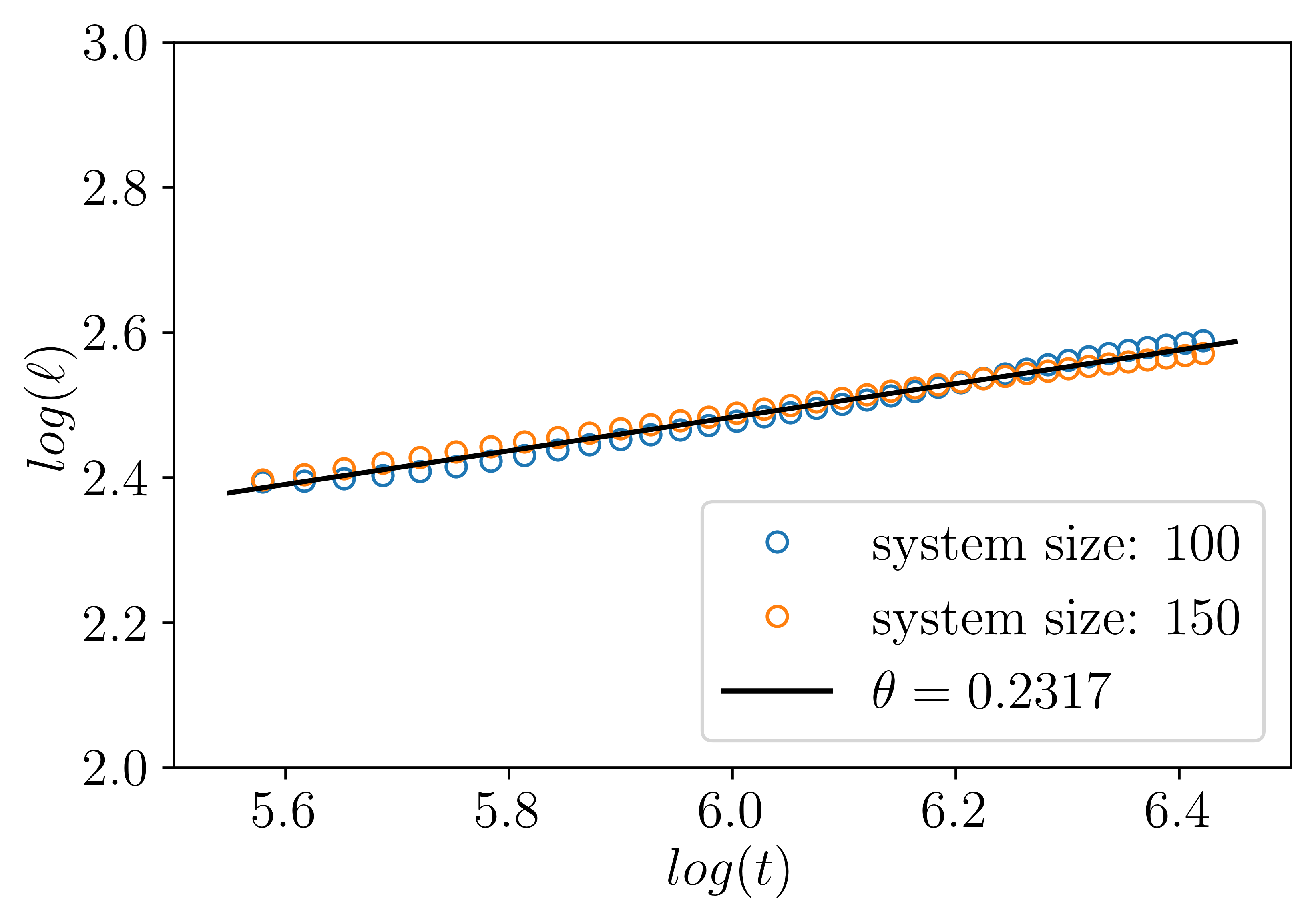} 
        \caption{Plot of the characteristic lengthscale $\ell$ as a function of time. Blue and orange circles are numerical result corresponding the systems of different size (minimal  model, $\Xi=0,\, \rho_0=1.1\rho_{cr}$); the solid line is the power-law fit $\ell\sim t^{\theta}$.}
\label{coarsening}
    \end{minipage}
\end{figure}

After a brief transient (not shown) we observe that the length scale $\ell$ of nematic domains grows as $\ell\sim t^{\theta}$, where $\theta\approx0.25$ (Fig.~\ref{coarsening}), in line with numerical results obtained for growth of passive nematic droplets [2]. We note that the value of the exponent is also numerically close to the one observed for the growth of droplets of spherical self-propelled particles in motility-induced phase separation~[3]. 

In the regime where we observe dry active turbulence, domains transiently coarsen to form one or few bands, however they undergo subsequent instabilities according to the mechanism described in the main text.

\if{
\section{Density flux correlations}

In the main text we showed the Fourier transform of $\ell(t)$ in Figure~2h in the case leading to dry active turbulence: this shows a power-law decay in frequency, which is often associated with chaotic or turbulent behaviour (although this is in not sufficient to rigorously prove that the dynamics is chaotic). 

Inspired by previous work on turbulence which characterised the decay of velocity-velocity correlations in wet active nematics~\cite{Thampi2013}, we also defined and computed correlations of the density flux ${\mathbf J}$ for our active turbulent patterns, as follows (note that there is no fluid velocity as our system is dry). 
First, we define the density flux ${\mathbf J}$ by writing our evolution equation for the density~\ref{SMMinModel1} as a conservation law, as follows,
\begin{equation}
	\partial_t\rho+\partial_iJ_i=0.
\end{equation}
Next, we calculate the flux correlation function (FCF) as $\langle \mathbf J(t,\mathbf r)\cdot\mathbf J(t,\mathbf r+\mathbf R) \rangle$ and its associated (spatial) Fourier transform, which we call $E(k)$ (in analogy with the power spectrum computed in studies of turbulence). 

Results show that the FCF intercepts $0$ for a distance $R_c$, roughly corresponding to the width of nematic bands (Figs.~\ref{correlations}\,a,b). Its Fourier transform displays a power-law decay, consistent with long-range spatial correlations and a size-dependent value for $R_c$ (Fig.~\ref{correlations}c). For completeness we also plot the structure factor, which also shows a power-law decay, consistent with a disordered lamellar phase. In the future, it would be of interest to see whether the behaviours shown in Figure~\ref{correlation} can be predicted analytically for our dry active turbulent pattern.

\begin{figure}[h!]
    \centering
    \begin{minipage}{\textwidth}
        \centering
        \includegraphics[width=0.6\textwidth]{together.png} 
        \caption{
        (a) Plot of the circularly averaged FCF $\langle \mathbf J(t,\mathbf r)\cdot\mathbf J(t,\mathbf r+\mathbf R) \rangle$  as a function of $R$.
        (b) Corresponding simulation snapshot, white segment illustrates the characteristic $R_c$, colorfunction is distribution of the MT density $\rho$.
        (c) Plot of the ``power spectrum'' $E(k)$, defined as the circularly averaged Fourier transform of the FCF.
        (d) Plot of the circularly averaged structure factor $S=\langle\rho(t,\mathbf k)\rho(t,-\mathbf k)\rangle$.
        (All plots refer to the minimal  model, $\Xi=0.3,\, \rho_0=1.1\rho_{cr}$, system size: $150\times150$).}
    \end{minipage}
\end{figure}
}\fi

\if{
\section{Turbulence regime}
\begin{figure}[h!]
    \centering
    \begin{minipage}{\textwidth}
        \centering
        \includegraphics[width=0.7\textwidth]{Chaosm43.png} 
        \caption{Time Fourier transform of the length scale $L(t)$ 
        ($\Xi=0.1,\, \rho_0=1.1\rho_{cr}$).}
    \end{minipage}
\end{figure}
}
\fi

\section{Captions for Supplementary Movies}

{\bf Suppl. Movie 1.} Movie showing the simulation results for the evolution of density and nematic ordering in the full model, with $\Xi=0$, and $\rho_0=20.735$. System size: $50\times 50$; $dx=0.5$, $dt=0.005$. The movie illustrates formation of a single steady state band in the case of isotropic interaction rates. \\

{\bf Suppl. Movie 2.} Movie showing the simulation results for the evolution of density and nematic ordering in the full model, with $\Xi=0.1$, $\tau_0=0.5$, and $\rho_0=19.747$. System size: $150\times 150$; $dx=0.5$, $dt=0.005$. The movie illustrates the dry active turbulence regime and corresponds to Figure 2b of the main text.\\

{\bf Suppl. Movie 3.} Movie showing the simulation results for the evolution of density and nematic ordering in the full model, with $\Xi=0.1$, $\tau_0=0.5$, $\rho_0=19.747$. System size: $50\times 50$; $dx=0.5$, $dt=0.005$. The movie illustrates the mechanism of band disruption and reformation in the dry active turbulence regime and corresponds to Figure 2c of the main text. \\

{\bf Suppl. Movie 4.} Movie showing the simulation results for the evolution of density and nematic ordering in the minimal model, with $\Xi=0$, $\rho_0=20.735$. System size: $50\times 50$; $dx=0.5$, $dt=0.005$. The movie illustrates formation of a single steady state band in the case of isotropic interaction rates in the minimal model. \\

{\bf Suppl. Movie 5.} Movie showing the simulation results for the evolution of density and nematic ordering in the minimal model, with $\Xi=0.1$, $\tau_0=0.5$, $\rho_0=19.747$. System size: $50\times 50$; $dx=0.5$, $dt=0.005$. The movie illustrates the dry active turbulence in the minimal model and corresponds to Figure 3a of the main text. \\

{\bf Suppl. Movie 6.} Movie showing the simulation results for the evolution of density and nematic ordering in the minimal model, with $\Xi=0.3$, $\tau_0=0.5$, $\rho_0=18.03$. System size: $150\times 150$; $dx=0.5$, $dt=0.005$. The movie illustrates the dry active turbulence in the minimal model in a large system and corresponds to Figure 3b of the main text. \\

{\bf Suppl. Movie 7.} 
Movie showing the simulations results for the evolution of density and nematic ordering in the modified minimal model in which $\beta_1$, $\beta_2$ , and $\zeta$ are equal to zero ($\Xi=0$, $\rho_0=20.735$; system size: $50\times 50$; $dx=0.5$, $dt=0.005$). The movie illustrates the mechanism of band formation and corresponds to Figure 3c of the main text. \\

{\bf Suppl. Movie 8.} Movie showing the simulations results for the evolution of density and nematic ordering in the modified minimal model in which $\chi$ is equal to zero ($\Xi=0$, $\rho_0=20.735$;
system size: $50\times 50$; $dx=0.5$, $dt=0.005$). The movie illustrates the role of $\chi$. \\

{\bf Suppl. Movie 9.} Movie showing the simulations results for the evolution of density and nematic ordering in the modified minimal model in which parameter $\zeta$ is equal to zero ($\Xi=0$, $\rho_0=20.735$;
system size: $50\times 50$; $dx=0.5$, $dt=0.005$). The movie illustrates the dry active turbulence regime with $\zeta=0$ and corresponds to Figure 3d of the main text. \\

{\bf Suppl. Movie 10.} 
Movie showing the simulations results for the evolution of density and nematic ordering in the modified minimal model in which $\beta_1$ and $\beta_2$ are equal to zero ($\Xi=0$, $\rho_0=20.735$; system size: $50\times 50$; $dx=0.5$, $dt=0.005$). The movie illustrates band undulation and corresponds to Figure 3e of the main text.\\

{\bf Suppl. Movie 11.} Movie showing the simulations results for the evolution of density and nematic ordering in the modified minimal model in which $\beta_1$ and $\beta_2$ are equal to zero while $\zeta$ is multiplied by $1.1$ ($\Xi=0$, $\rho_0=20.735$; system size: $50\times 50$; $dx=0.5$, $dt=0.005$). The movie illustrates a pathway to chaotic dynamics based on band undulations. \\

{\bf Suppl. Movie 12.} Movie showing the simulations results for the evolution of density and nematic ordering in the modified minimal model in which $\beta_1$ and $\beta_2$ are equal to zero while $\zeta$ is multiplied by $1.5$ ($\Xi=0$, $\rho_0=20.735$; system size: $50\times 50$; $dx=0.5$, $dt=0.005$). The movie illustrates a similar pathway to chaotic dynamics as in Suppl. Movie 11.

\noindent
{\small
\\[2mm]
$[1]$ I. Maryshev, D. Marenduzzo, A. B. Goryachev, and A. Morozov, Phys. Rev. E \textbf{97}, 022412 (2018).\\[0.5mm]
$[2]$ M. Mata, C. J. Garc{\'\i}a-Cervera, and H. D. Ceniceros, J. Non-Newtonian Fluid Mech. \textbf{212}, 18 (2014).\\[0.5mm]
$[3]$ J. Stenhammar, A. Tiribocchi, R. J. Allen, D. Marenduzzo, and M. E. Cates, Phys. Rev. Lett. \textbf{111}, 145702 (2013).
}

\end{document}